\documentclass[prl,twocolumn,superscriptaddress,nopacs,amsmath,amssymb,nofootinbib]{revtex4-2}

\usepackage{graphicx}
\usepackage{hyperref}
\usepackage{color}
\usepackage{booktabs}
\usepackage{makecell}
\usepackage{soul}

\newcommand{\YT}{\textcolor{black}}

\usepackage{verbatim}

\usepackage{lineno}

\begin{document}

\title{Doping a Wigner-Mott insulator: \\ Exotic charge orders in transition-metal dichalcogenide moir\'e heterobilayers}
\author{Yuting Tan}
\affiliation{Department of Physics and National High Magnetic Field Laboratory, Florida State University, Tallahassee, FL, USA}
\affiliation{Condensed Matter Theory Center and Joint Quantum Institute, Department of Physics, University of Maryland, College Park, Maryland 20742, USA}
\author{Pak Ki Henry Tsang}
\affiliation{Department of Physics and National High Magnetic Field Laboratory, Florida State University, Tallahassee, FL, USA}
\author{Vladimir Dobrosavljevi\'{c}}
\affiliation{Department of Physics and National High Magnetic Field Laboratory, Florida State University, Tallahassee, FL, USA}
\author{Louk Rademaker}
\affiliation{Department of Theoretical Physics, University of Geneva, 1211 Geneva, Switzerland}
\begin{abstract}
The moir\'{e} pattern induced by lattice mismatch in transition-metal dichalcogenide heterobilayers causes the formation of flat bands, where interactions dominate the kinetic energy. At fractional fillings of the flat valence band, the long-range electron interactions then induce Wigner-Mott crystals. In this Letter we investigate the nontrivial electronic phases appearing away from fractional fillings. Here, competing phases arise that are either characterized as doped Wigner-Mott charge transfer insulators, or alternatively, a novel state with frozen charge order yet is conducting: the `electron slush'. We propose that an extremely spatially inhomogeneous local density of states can serve as a key signature of the electron slush.
\end{abstract}

\maketitle

When two atomically thin materials are intentionally misaligned or have a lattice mismatch, a long-range geometric moir\'{e} pattern emerges. This results in a drastic reduction of the electronic kinetic energy,  paving the way for new strongly correlated phases.\cite{Balents:2020.725B,Andrei:2020asg} 
Though the first signs of correlations were seen in twisted bilayer graphene \cite{Cao:2018ff,Cao:2018kn},
it is particularly appealing to use instead monolayer semiconductor transition metal dichalcogenides (TMDs) MX$_2$ where M = W, Mo and X = S, Se, Te.\cite{Mak:2010do,Liu:2013bua,Geim:2013hf,Manzeli:2017tmd,Mak:2022review,Giacomoarxiv2022}
In TMD moir\'{e} bilayers, the effective interaction strength $U$ scales as the inverse moir\'{e} length $U \sim a_M^{-1}$ whereas the flat band kinetic energy at the top of the valence band scales as $W \sim a_M^{-2}$.  
Consequently, this back-of-the-envelope argument suggests $U/W \sim a_M$, meaning there is no limit as to how strongly coupled a TMD bilayers can be (Appendix A)!

\begin{figure}[t!]
	\includegraphics[height=8cm]{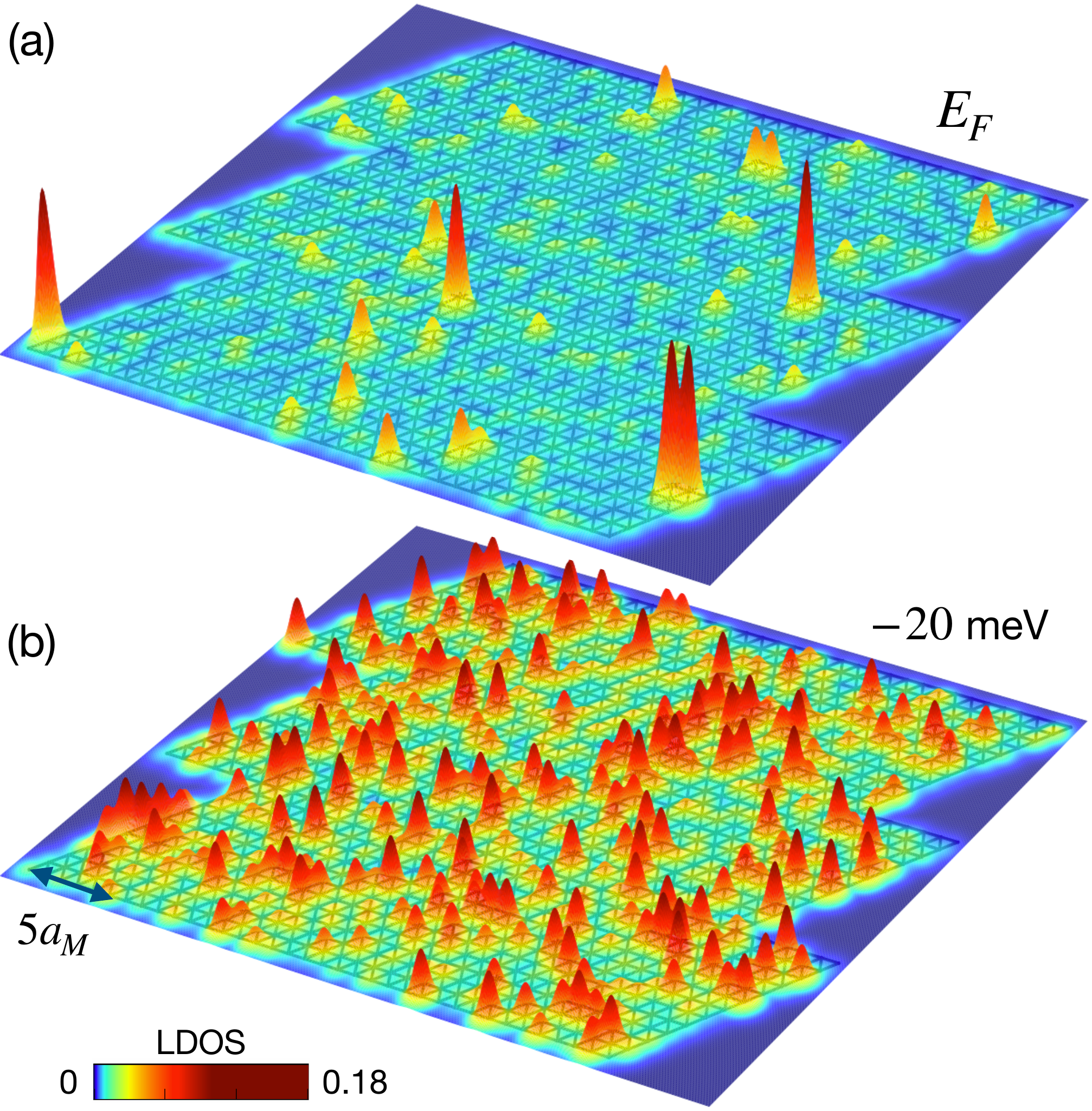}
	\caption{The local density of states (LDOS) of a typical ``electron slush'' phase at incommensurate fillings of a TMD heterobilayer, obtained using \YT{non-charge-self-consistent Hartree+CPA+DMFT}, with $T=1.2$K, $n=1/2$, and $U/t=38.3$. Here, the electrons are frozen in an amorphous configuration. The presence of screened Coulomb interactions causes a soft spectral gap (Fig.~\ref{Fig:DOS}) and nonzero conduction (Fig.~\ref{Fig:Transport}). However, the LDOS is spatially extremely inhomogeneous, and dominated by only a few peaks at the Fermi level (a). At lower energies, here at $E=-20$ meV (b), the LDOS shows more structure of the amorphous slush state, which can be observed using scanning tunneling microscopy. }
	\label{Fig:Amorphous}
\end{figure}

Furthermore, combining various TMD monolayers with or without a relative twist angle allows for a wide degree of variability in constructing the physics of flat bands \cite{Carr:2020cm}. Most notably, in TMD {\em hetero-bilayers} (combining two different TMDs) the electronic states are localized in only one layer and form an effective triangular lattice Hubbard model \cite{Wu:2018ic,Zhang:2019wk,Zhang:2021ix,Li:2021cc,Vitale:2021vg,Rademaker:2021arXivSOC}.

The observation of a Mott insulator state at half-filling $n=1$ of the first flat valence band confirmed the presence of strong electronic correlations in aligned WSe$_2$/WS$_2$ \cite{Tang:2020bb,Li:2021cd,Li:2021vy} .
Subsequently a range of charge ordered generalized Wigner-Mott crystals \cite{Camjayi:2008jh,Amaricci:2010dw,Radonjic:2012gm} at fractional fillings $n=1/2, 1/3, 2/3, 1/4,$ and so forth, was observed; meaning not only the onsite repulsion but also the longer ranged Coulomb interaction plays a central role \cite{Regan:2020fk,Xu:2020dx,Huang:2021io,Jin:2021es,LiWang:2021stmWigner,LiWang:2021stmLocal,Miao:2021exciton,Liu:2021exciton,Li:2021capacitance}.

Exactly at these fractional fillings, the Wigner-Mott crystal is an insulator, which can be described on the mean field level \cite{Zhang:2019wk,Pan:2020cd,Padhi:2021bs,Zhang:2021hx,Pan:2020kga,Musser:2021arXiv}. Away from fractional fillings, the situation becomes less clear. While there is a large theoretical literature on interacting triangular lattice models, most focus on superconducting instabilities\cite{Nandkishore.2014,Venderley.2019,Wolf.2022,Huang.2022vn,Gneist.2022}, which is not observed in experimental moir\'e systems. Instead, a realistic possibility is the emergence of competing charge ordered phases, as was recently suggested by Monte Carlo simulations \cite{Matty:2021wignermott}. Alternatively, the system can become a {\em doped} Wigner-Mott insulator \cite{Lee:2006de} where the charge order is given by the nearest fractional filling. 

\begin{figure*}[t]
        \includegraphics[width=\textwidth]{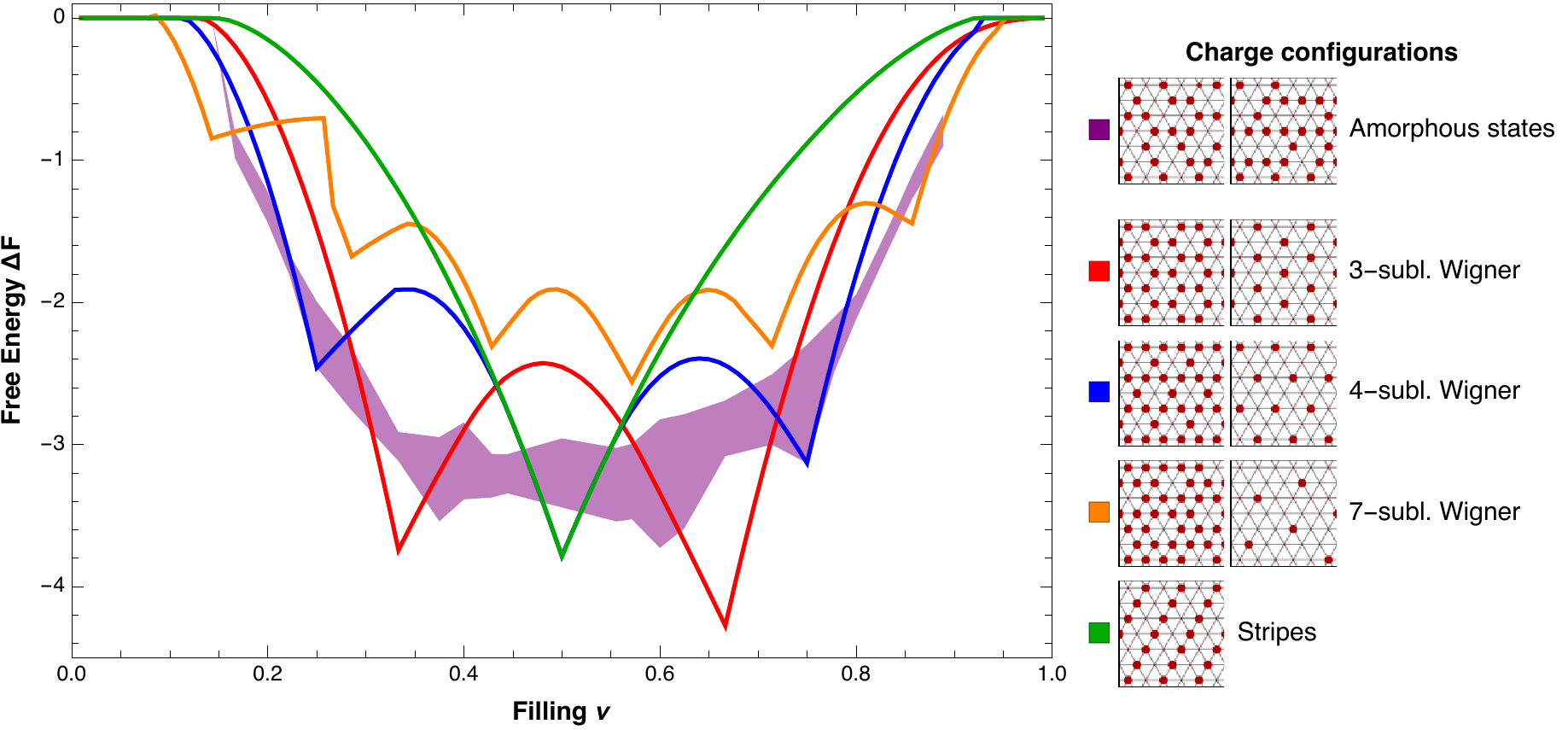}
	\caption{Results of the self-consistent Hartree calculation for charge ordering with $V_0 = 30$ meV. We show the net free energy gain for various charge patterns, including the stripe phase, and generalized Wigner crystal phases at $n =1/3, 1/4$ and $1/7$. On large $12\times12$ unit cells we also considered the possibility of amorphous charge states, which have the lowest free energy in some ranges of filling.}
	\label{Fig:Wigner}
\end{figure*}

Aside from variational approaches close to integer filling\cite{Laubach.2015}, studies of the interplay of Mottness and charge order beyond mean-field theory are all based on versions of DMFT, augmented with either self-consistent Hartree-Fock for the charge order\cite{Camjayi:2008jh,Amaricci:2010dw,Radonjic:2012gm,Merino:2013dk,Ralko:2015df} or GW for the momentum-dependent self-energy.\cite{Hansmann.2013,Chen.2021m1} Following the former approach, at most incommensurate fillings we identify a novel metastable {\em "electron slush"} phase. Here the electron charge freezes {\em spontaneously} in amorphous real-space patterns, whose signatures in the local density of states are shown in Fig.~\ref{Fig:Amorphous}.
Despite the frozen charge order, the screening of long-range interactions prevents the opening of a spectral “Coulomb" gap. In addition, the random electrostatic fields produced by such amorphous charge order induce local {\em doping} of the Wigner-Mott solid, resembling the effect of site disorder in ordinary Mott insulators.

As a result, the system remains a conductor  — albeit a very bad one, with an extremely high resistivity surpassing the Mott-Ioffe-Regel limit.  This stands in great contrast to competing “doped Mott-Wigner insulator" phases, where the fractional periodic charge order remains, and the dopants realize a strongly  renormalized (but highly conducting!) Fermi liquid at low temperatures. Physically, this amorphous charge order can be seen as a "compromise" between several closely competing periodic Wigner orders, at incommensurate fillings. The strange spectral and transport properties characterizing such amorphous Mott matter are the central discovery of this work. \YT{We emphasize that the frozen charge order is the result of very strong nonlocal Coulomb repulsion $V/t \gg 1$, which can be captured self-consistently on a semi-classical Hartree level. On top of that, we use DMFT to gain insights into the effect of such charge freezing upon the spin and transport effects associated with strong electronic correlations due to the Hubbard $U$.}


\begin{figure*}[t]
	\includegraphics[width=0.8\textwidth]{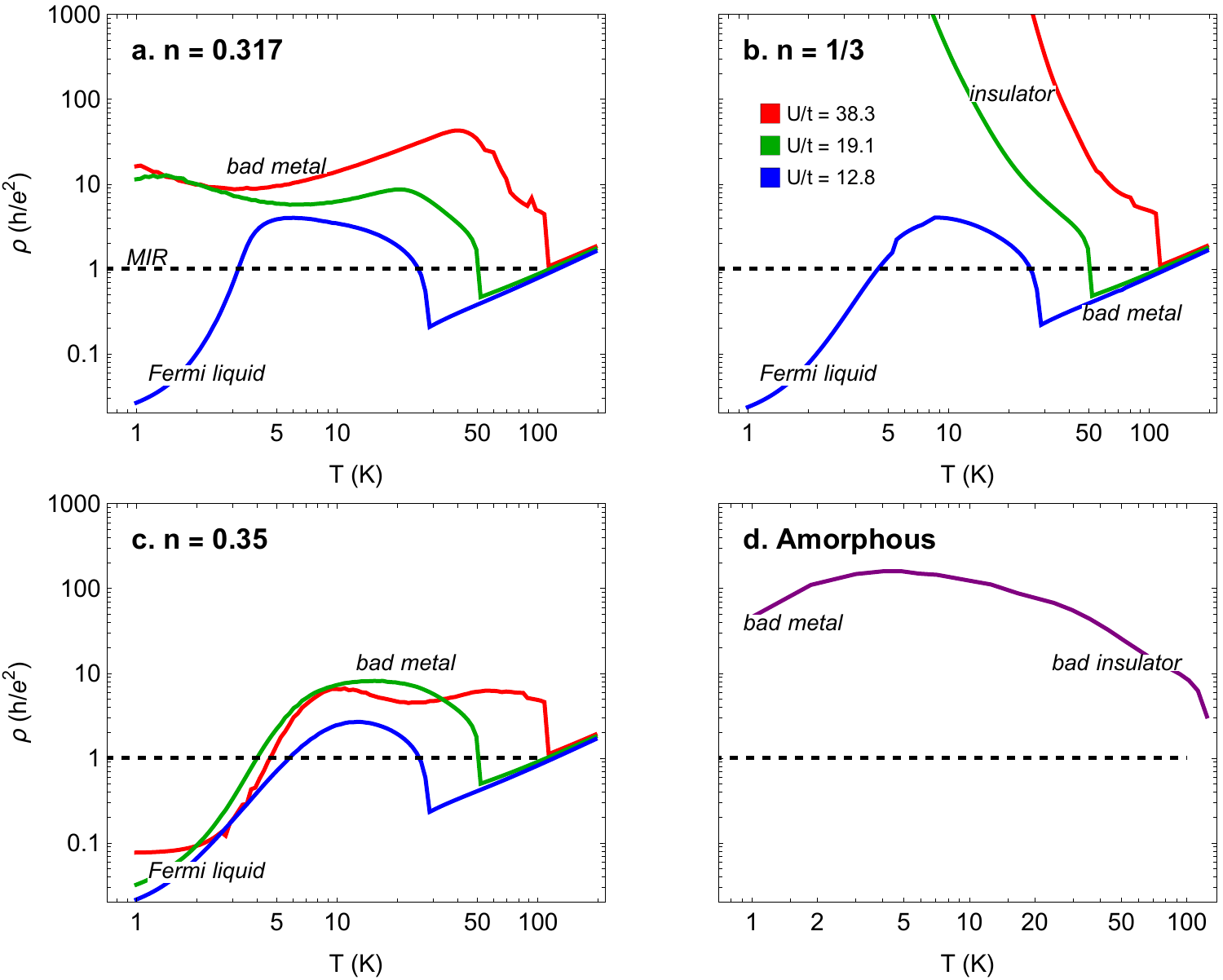}
	\caption{DC resistivity in the vicinity of the $n=1/3$ moir\'{e}-Wigner-Mott insulator ({\bf a}-{\bf c}) and for amorphous states close to $n=1/2$ ({\bf d}). Exactly at $n=1/3$ ({\bf a}) we find a strange metal phase at high temperatures, followed by activated behavior after charge-order sets in. For weak interactions, the Hubbard-Mott gap closes, and a Fermi liquid forms at low temperature. Heavy Fermi liquid behavior remains the dominant low-temperature behavior upon electron-doping ({\bf c}); for hole-doping ({\bf a}), Fermi liquid condensation occurs only at extremely low temperatures (not shown). In contrast, the amorphous state ({\bf d}, shown for $U/t=38.3$, at $n = 1/2$) is characterized by a weakly temperature-dependent resistivity turning from insulating to bad metallic behavior. The dashed line indicates the Mott-Ioffe-Regel limit.}
	\label{Fig:Transport}
\end{figure*}


{\em Moir\'{e} model --} To model the aligned WS$_2$/WSe$_2$ hetero-bilayers we follow the continuum formulation proposed by Wu et al. \cite{Wu:2018ic}. We briefly summarize their approach here, a more detailed description is provided in Appendix A.  The continuum model starts with the top of the valence band in monolayer WSe$_2$, where the holes have an effective mass of $m^* = 0.36 m_e$.

Aligned ($\theta = 0$) WS$_2$/WSe$_2$ heterobilayers have a moir\'{e} length of approximately $a_M = 7.9$ nm. To obtain the strength and phase of the periodic moir\'{e} potential, we performed density functional theory calculations using the approach of Refs.~\cite{Wu:2018ic,2018PhRvB..97c5306W,2017PhRvL.118n7401W}. For WS$_2$/WSe$_2$ bilayers we obtained $(V,\psi) = (7.7$ meV$, -106^\circ$). Note however, that the accuracy of the {\em magnitude} $V$ is debatable. Recently, STM measurements revealed a moir\'{e} potential in excess of 100 meV \cite{Shabani:2021fh}, whereas theoretical work suggests a complete absence of a moir\'{e} potential \cite{Vitale:2021vg}. Additionally, the magnitude is very sensitive to the precise interlayer distance (and can thus be tuned using pressure as was done in twisted bilayer graphene \cite{Yankowitz:2018tx}). Therefore, one should bear in mind that the provided values for the electronic bandwidth can be significantly different in real systems.

Nevertheless, with this value of moir\'{e} potential a flat band emerges that can be represented by the extended Hubbard model on a triangular lattice. The nearest neighbor hopping is $t = 1.9$ meV, and we ignore longer-ranged hoppings. The corresponding Wannier orbitals \cite{Marzari:2012eu} are approximately Gaussian around the W/Se stacking center with width $\sigma = 0.125 a_M$. We use the  screened Coulomb potential $V_{\mathrm{sc}}(r) = \frac{e^2}{4 \pi \epsilon}\left( \frac{1}{r} - \frac{1}{\sqrt{r^2+ d^2}}\right)$ to calculate the effective interaction strengths. With a screening length of $d=20$ nm, this provides an onsite repulsion of $\epsilon U = 1.7$ eV, and a nearest neighbor repulsion of $\epsilon V = 0.6$ eV. Even with quite large values for the dielectric constant the onsite repulsion is stronger than the bandwidth, and indeed even $V \gg t$!

Because both the bandwidth (via the moir\'{e} potential and interlayer coupling) and the interaction strength (via engineering of the dielectric environment) are tunable, in the remainder of this paper we vary the ratio $U/t$. Our choices are $U/t = 38.3, 19.1$ and $12.8$, corresponding roughly to dielectric constants of $\epsilon = 23.4, 46.8$ and $70.2$, respectively. For the strongest choice of $U/t$, the corresponding nearest neighbor repulsion becomes $V/t = 13$ -- large enough to induce nontrivial charge order. 
%

{\em Charge order --} Having established that the long-range repulsion is quite strong, we next discuss the development of charge order. In the absence of a lattice potential, a dilute two-dimensional electron gas will form a triangular {\em Wigner crystal} \cite{kravchenko2017strongly,Tan:2022crystals}. Whenever an underlying triangular lattice is present, the triangular Wigner crystal can only be formed at fractional fillings given by $n = \frac{1}{m_1^2+m_1m_2+m_2^2}$ with $m_i$ integers. The largest fractions are $n = 1/3, 1/4, 1/7,$ and $1/9$; one can also have a generalized Wigner crystal of {\em holes} at fillings $n = 2/3, 3/4$, etc.
At filling $n = 1/2$ the long-range repulsion favors stripe order \cite{Mahmoudian:2014vh}. At all of these fractional fillings a correlated insulator state has been observed in WSe$_2$/WS$_2$ \cite{Regan:2020fk,Xu:2020dx,Huang:2021io,Jin:2021es}.

Because the onsite repulsion $U$ is large, to leading order we can project out double occupancy. As a first step to identify various forms of charge order across the phase diagram, we use self-consistent Hartree theory for long-range interacting {\em spinless} electrons, with screened Coulomb interaction $V({\bf r}) = \frac{V_0}{r} e^{(r-1)/d} $ where $r$ is measured in moir\'{e} lattice constants, $d = 20$ nm the screening length and we choose $V_0 = 30$ meV (Appendix B). The zero-temperature free energy of various charge ordering patterns is shown in Fig.~\ref{Fig:Wigner}, left. Here we compare the free energy of five different possible charge patterns as a function of filling $n$. The most robust of these configurations is the Wigner-Mott crystal with 3 sublattice sites, yielding a triangular Wigner crystal at $n = 1/3$ and a honeycomb crystal at $n = 2/3$. Similarly structures appear at $n = 1/4$, $1/7$, and other fractional fillings. In addition, we confirm the presence of the $n = 1/2$ stripe phase \cite{Mahmoudian:2014vh,Jin:2021es}.

Throughout the entire range of filling $0 < n < 1$, we further compare these Wigner crystal states with aperiodic frozen charge patterns, also known as {\em amorphous} states. For this we used a large 12$\times$12 unit cell, equivalent to 144 moir\'{e} unit cells, and found a local minimum of the Hartree free energy corresponding to an inhomogeneous charge distribution.  
While most electron crystal states are stable upon small doping, 
there appears {\em in between each rational fillings} a regime where amorphous patterns are the most stable ones. If the system is cooled fast enough and with infinitesimal disorder, however, crystalline orders or macroscopic phase separation can be avoided at {\em all} fillings. The electrons then freeze into an amorphous solid, behavior that has been observed in $\theta$-organic materials \cite{Kagawa:2017de,Sato:2014cl,Kagawa:2013hz}, consistent with recent theory \cite{Mahmoudian:2014vh}. The emergence of similar metastable amorphous structures has also been reported in various TMD layered materials featuring Wigner crystals and glasses made of "star of David" polarons\cite{gerasimenko2019intertwined,gerasimenko2019quantum}.

The results presented in Fig.~\ref{Fig:Wigner} were obtained for the interaction strength $V/W \approx 2$ -- nearest neighbor repulsion is about twice the noninteracting bandwidth. This ratio is highly dependent on the dielectric environment and the strength of the moir\'{e} potential. With the parameters used here we find that the $n=1/3$ Wigner crystal is stable up to $T_c \approx 12$ meV, which is a factor 2-3 higher than observed in experiments \cite{Regan:2020fk,Huang:2021io}. It is therefore likely that in real materials $V/t \approx 5$, which would correspond to $U/t \approx 15$. Since mean field theory systematically overestimates the tendency to order, these approximations for $U/t$ and $V/t$ are lower bounds.

\begin{figure*}[t]
	\includegraphics[width=0.8\textwidth]{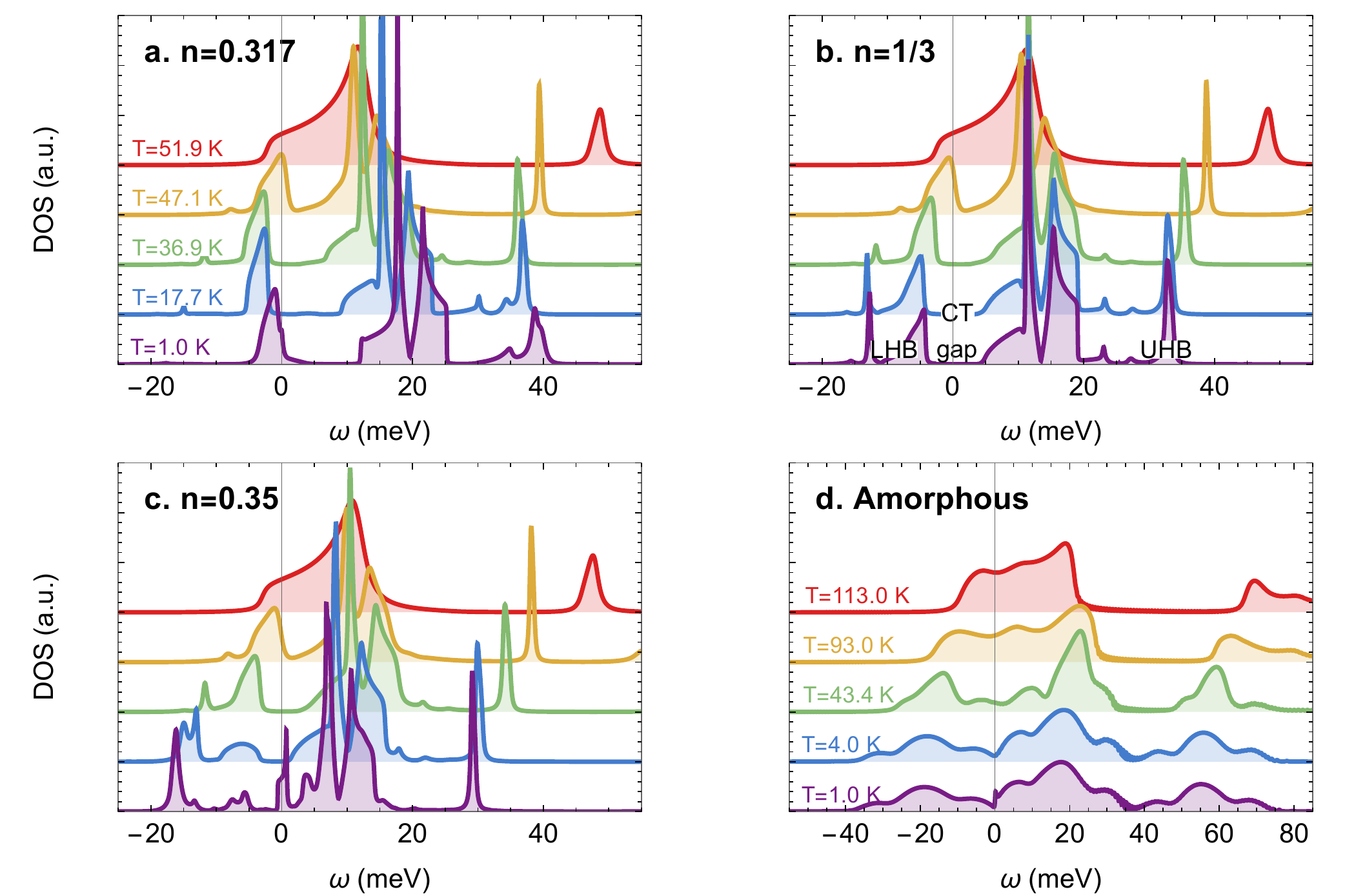}
	\caption{The density of states (DOS) as a function of energy for various doping and temperature. We only show here data for $U/t = 19.1$. At $n=1/3$ ({\bf b}), the spectrum is clearly split into a lower Hubbard band (LHB), a middle band consisting of the states on the unoccupied honeycomb sublattice, and the upper Hubbard band (UHB). The gap at the Fermi level is therefore a charge-transfer (CT) gap. Consequently, electron-doping ({\bf c}) leads to the formation of a quasiparticle peak at the Fermi level whereas hole-doping ({\bf a}) puts the Fermi level in the incoherent lower Hubbard band. Bad metal behavior, as shown in Fig.~\ref{Fig:Transport}, is due to a shifting chemical potential which can be clearly seen in the left figure. Upon increasing temperature the charge-transfer gap closes and a strange metal phase appears. The DOS for the amorphous state ({\bf d}) has similar features (LHB, CT gap, UHB) but its features are smeared out due to the inhomogeneous charge patterns, leading to a soft gap in the spectrum and consequently bad metal behavior (Fig.~\ref{Fig:Transport}).}
	\label{Fig:DOS}
\end{figure*}

{\em Doped Wigner-Mott insulators --}
\label{Sec:Mott}
Having established various forms of charge order present, we now investigate their impact on Mott-Hubbard correlations, within a setup based on dynamical mean-field theory (DMFT)\cite{Georges:1996un,Zang2022PRX};
a detailed description of our calculation methods can be found in Appendix C. Representative results for the resistivity as a function of temperature are shown in Fig.~\ref{Fig:Transport}, and the corresponding spectral functions are shown in Fig.~\ref{Fig:DOS}.

Our starting point is the Wigner-Mott insulator, such as the state we find at $n=1/3$ filling. In our system, the on-site repulsion $U$ is generally larger than the gap induced by the charge order, so the resulting Wigner-Mott state can therefore be recognized as a {\em charge-transfer} insulator \cite{Zaanen:1985bsa,Rademaker:2018fy,Zhang:2019wk}. Here the electronic states below the Fermi level are Mott-localized on the occupied sublattice, whereas the states above the Fermi level reside on the unoccupied honeycomb-like sublattice. This is illustrated in Fig.~\ref{Fig:DOS}b, where for $n = 1/3$ we indicate the charge transfer (CT) gap, and the upper and lower Hubbard band (UHB/LHB).

Exactly at fractional filling, the strong onsite repulsion leads to three distinct transport regimes, shown in Fig.~\ref{Fig:Transport}b. At high temperatures above the crystalline ordering transition $T>T_c$ we find a typical $\rho = A + BT$ linear resistivity {\em bad metal} regime with resistivity comparable or even above the Mott-Ioffe-Regel limit.
Just below the charge-ordering transition ($T\leq T_c$) the resistivity jumps up and the system becomes insulating. When the interactions are weaker, in our model for $U/t = 12.8$, we find an intermediate situation where the charge order can form but the onsite $U$ is not strong enough to open a Mott gap, leading to low $T$ Fermi-liquid behavior with well-defined quasiparticles. Note that this is in contrast to Ref.~\cite{Musser:2021arXiv} where it is {\em assumed} that the Mott and the charge order transition always happen simultaneously. In our picture, as in previous work on Wigner-Mott systems \cite{Camjayi:2008jh,Amaricci:2010dw,Radonjic:2012gm}, the weakly first-order Mott-like metal-insulator transition occurs {\em within} the charge-order state. Our scenario is actually realized in many real materials \cite{gerasimenko2019intertwined,gerasimenko2019quantum} featuring Wigner crystallization in lattice systems, which can lead to either insulating {\rm or} metallic charge-ordered ground states.

Let us now turn our attention to {\em doping} the Wigner-Mott state. For weaker interactions ($U/t = 12.8$), doping a metallic charge density wave retains the conventional Fermi liquid behavior. However, when the onsite Mott correlations are large we see different behavior, depending on whether we dope with holes or electrons.

For {\em electron-doping} ($n = 1/3 + \delta$ with $\delta >0$, Figs.~\ref{Fig:Transport}c and \ref{Fig:DOS}c), the dopants appear in an otherwise empty honeycomb sublattice. As a result, the dopants are moderately interacting and form {\em pinball liquid} \cite{Ralko:2015df,Merino:2013dk} with a renormalized Fermi liquid regime at low temperature. We find that the associated Fermi liquid coherence temperature vanishes {\em linearly} as we approach Wigner-Mott insulator, $T_F \sim |n - n_c|$. This is a clear prediction of our theory that can be tested in experiment, similar to recent measurements of bandwidth-controlled Mott criticality \cite{Li:2021cd,Ghiotto:2021qc}.

This Fermi liquid regime evolves at intermediate temperature into an incoherent high-resistivity phase. Here, the shifting of the chemical potential into the charge ordering gap competes with the loss of quasiparticle coherence, resulting in non-monotonic behavior of the resistivity. Finally, as is the case at fractional filling, when charge order is lost the high-temperatures, linear-$T$ bad metal behavior is recovered.

Upon {\em hole-doping} ($n = 1/3 - \delta$, Figs.~\ref{Fig:Transport}a and \ref{Fig:DOS}a), the dopants are populating the sites of the triangular Wigner-Mott lattice. While at extremely low temperatures a FL regime does appears (not shown), for most temperatures the Fermi level lies within the incoherent lower Hubbard band, leading to bad metal behavior. As the temperature increases, the chemical potential shifts into the charge-transfer gap leading to alternating weakly insulating and weakly metallic states. More details of the transport behavior within the entire regime of dopings and temperatures around the Wigner-Mott regime are shown in Appendix D.


{\em Electron slush --}
\label{Sec:Slush}
As shown in Fig.~\ref{Fig:Wigner}, however, for the largest part of the phase diagram, at most incommensurate fillings, the system is likely to freeze in an amorphous charge pattern. Because any amorphous charge order results in a random distribution of internal electrostatic potentials, within our DMFT setup the rest of the calculation reduces to solving an appropriate Hubbard-like (i.e. charge-transfer) model supplemented with random site energies, which we tackle within the well-known CPA-DMFT approach \cite{aguiar2005prb} (see Appendix C and E).
In the following we focus on the amorphous states at $n = 1/2$ and for $U/t = 19.1$, shown in Figs.~\ref{Fig:Transport}d and \ref{Fig:DOS}d; within our setup, the qualitative behavior at other fillings and other (comparable) values of $U/t$ is essentially the same. Note that $n=1/2$ is a fractional and not an incommensurate filling; we performed our DMFT calculations at this filling for convenience; the properties of the electron slush are the same at fractional and incommensurate fillings. 

Even though the electrons clearly develop local frozen charge order, this does not lead to the opening of a hard gap in the spectrum. This situation reminds us of the formation of a soft “Coulomb" gap in  electron glasses \cite{1975JPhC....8L..49E,1976JPhC....9.2021E,Pramudya:2011iv,Rademaker:2018ku}. However, here the screening of the long-range Coulomb interaction (due to the presence of gate on the TMD heterobilayer) causes a further "filling" of the soft gap. As a result, despite the frozen charge order, there is no gap towards transport: the amorphous ordered state can still conduct, although they do so very poorly. Our DMFT results indicate at low temperatures the formation of weakly-developed quasiparticles around the Fermi level, with spectral features reflecting the amorphous charge background in question. This combination of frozen charge order with motion of electrons reminds us of the popular “slush” drinks which are liquids (conducting) yet frozen. We therefore propose to call this novel phase an {\em electron slush}. 

Note that the amorphous charge order of the electron slush can be characterized by the presence of short-range stripe correlations, which has also been observed in Monte Carlo simulations \cite{Mahmoudian:2014vh,Huang:2021io,Matty:2021wignermott}. Therefore, scanning tunneling microscopy (STM) \cite{LiWang:2021stmWigner,LiWang:2021stmLocal} {\em topography} should reveal the frozen charge configurations. A measurement of the {\em local density of states (LDOS)}, however, leads to an interesting local realization of the soft gap feature. As shown in Fig.~\ref{Fig:Amorphous}, the LDOS at the Fermi level is dominated by a few localized peaks at large distances, whereas at energies away from the Fermi level the amorphous structure is more spread out. This stark difference between topography and LDOS is another predicted feature of the electron slush phase, which awaits experimental verification. 


{\em Outlook --}
\label{Sec:Outlook}
Summarizing, we show that, for most fillings, an amorphous charge ordered metallic state -- the electron slush -- appears in strongly correlated TMD hetero-bilayers.
Our results suggest to perform transport and STM experiments on TMD heterobilayers to study the interplay between doping a Wigner-Mott insulator and amorphous charge order. 

Note that in this Letter we ignored the role of spin-orbit coupling (SOC) in the moir\'{e} flat bands \cite{Rademaker:2021arXivSOC}. Also, in this work we did not include the role of quenched disorder. It is known that TMD mono- and bilayers have significantly more disorder than other Van der Waals materials such as graphene, notably in the form of vacancies.\cite{Pasupathy:2022aspen} Since disorder strongly affects the zero-temperature limit of the resistivity, our transport results only apply at intermediate to high temperature. On the other hand, disorder is likely to further stabilize the electron slush phase since it promotes amorphous charge configurations. Our work shed light on the interesting but mysterious amorphous phase, which could very well play a central role in strongly correlated TMD heterobilayers at low band filling. Here we took the first steps to theoretically describe the likely features of this exotic regime, although more accurate treatments of the interplay between strong correlations and {\em induced/effective} disorder effects may be necessary, especially at the lowest temperatures. This research direction remains a challenge for future work.



\section{Acknowledgements}  We acknowledge discussions with Allan MacDonald, Fengcheng Wu, Jie Shan, Kin Fai Mak, Dima Abanin and Johannes Motruk.
L.R. was funded by the SNSF via Ambizione grant PZ00P2\_174208. Work in Florida (Y. T., P. K. H. T., and V. D.) was supported by the NSF Grant No. 1822258, and the National
High Magnetic Field Laboratory through the NSF Cooperative Agreement No. 1644779 and the State of Florida.

\section{APPENDIX A: Model parameters }
\label{Sec:Moire}

\begin{figure*}[ht]
	\includegraphics[width=\textwidth]{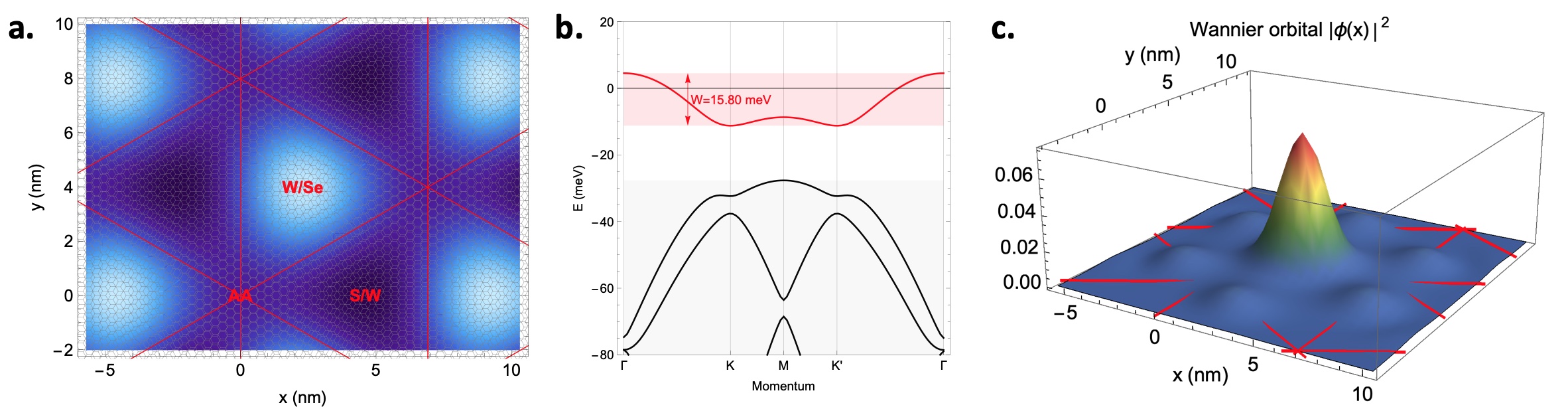}
	\caption{{\bf a.} The Moir\'{e} potential in aligned WS$_2$/WSe$_2$ heterobilayers, based on the continuum model from Ref.~\cite{Wu:2018ic} using our {\em ab initio} density functional calculations. There is a clear maximum at the W/Se$_2$ stacking in the Moir\'{e} unit cell.
	{\bf b.} The resulting bandstructure contains a flat band with bandwidth $W = 15.8$ meV separated from the other bands. 
	{\bf c.} The Wannier orbital corresponding to the flat band is approximately a Gaussian centered at the maximum of the Moir\'{e} potential.}
	\label{Fig:Moire}
\end{figure*}
A particularly insightful way to understand the flat bands in heterobilayers is the continuum model proposed by Wu et al.\cite{Wu:2018ic}. Here we apply their method to WS$_2$/WSe$_2$ heterobilayers.

In a single TMD monolayer the top of the valence band has a parabolic dispersion around the ${\bf K}$ and ${\bf K'}$ point. Strong spin-orbit coupling causes a spin-valley splitting, so that the states at ${\bf K}$/${\bf K'}$ are only singly degenerate and carry opposite spin. The band-structure in a single valley can thus be approximated as
\begin{equation}
	\epsilon ({\bf q}) = - \frac{\hbar^2 |{\bf q}|^2}{2 m^*}
	\label{Eq:ValenceBandTop}
\end{equation}
where ${\bf q} = {\bf k} - {\bf K}/{\bf K'}$ and the effective mass $m^* = 0.36 m_0$ for WSe$_2$.

Let us now look at a bilayer system. A monolayer WSe$_2$ has a lattice constant of approximately $a_{1} = 3.325$ \AA, whereas monolayer WS$_2$ has $a_{2} = 3.191$ \AA. Therefore even when the two layers are aligned, a moir\'{e} pattern emerges due to the lattice mismatch. The length scale of this pattern is given by 
\begin{equation}
	\frac{1}{a_M} = \sqrt{ \frac{1}{a_1^2} + \frac{1}{a_2^2} - \frac{2 \cos \theta}{a_1 a_2}} 
	\label{Eq:TwistedHeteroMoireLength}
\end{equation}
where $\theta$ is a twist angle. When the twist angle is small, and writing $a_2 \equiv a_1 (1 - \delta)$, we obtain
\begin{equation}
	a_M = a_1 / \sqrt{\delta^2 + \theta^2}.
\end{equation}
An {\em aligned} ($\theta = 0$) WS$_2$/WSe$_2$ heterobilayer therefore has a moir\'{e} length of approximately $a_M = 7.9$ nm. Note that only heterobilayers where the chalcogenides are the different in the two layers lead to a Moir\'{e} pattern when aligned. Furthermore, the Moir\'{e} length is maximal for our combination of S and Se: WS$_2$/WTe$_2$ has $a_M = 3.2$ nm whereas WSe$_2$/WTe$_2$ has $a_M = 5.0$ nm. This makes the platform WS$_2$/WSe$_2$ likely the most correlated among all aligned TMD heterobilayers.

Electronically, monolayer WS$_2$ has a larger bandgap than monolayer WSe$_2$, and the bilayer will have a type II band alignment of the two layers. This means that the top of the valence band of the bilayer consists of electronic states confined to the WSe$_2$ layer only. Note that a perpendicular electric field can change the band alignment of the two layers, thus bringing the WS$_2$ valence band at the same energy as the WSe$_2$ valence band, as was done recently in MoTe$_2$/WSe$_2$ bilayers.\cite{Li:2021vy} Here, however, we study the WS$_2$/WSe$_2$ bilayer in the absence of an electric field. In that case, the electronic states in the WSe$_2$ feel a position-dependent {\em Moir\'{e} potential} $\Delta({\bf r})$ due to the presence of the WS$_2$ layer.

The Moir\'{e} potential can be approximated using plane waves
\begin{equation}
	\Delta ({\bf r}) = \sum_{{\bf G}^{M}_i} V_i e^{i {\bf G}^{M}_i {\bf r}}
	\label{Eq:MoirePotential}
\end{equation}
where ${\bf G}^{M}_i$ with $i = 1 \ldots 6$ are the six reciprocal Moir\'{e} vectors. Without loss of generality, we set ${\bf G}^M_1 = \frac{4 \pi}{3 a_M} (1,0)$ and the other five are just sixfold rotations of the first reciprocal vector. Because $\Delta ({\bf r})$ must be real, and it has three-fold rotational symmetry, we have $V_1 = V_2 = V_5$ and $V_1 = V_4^*$,\cite{Wu:2018ic,2018PhRvB..97c5306W,2017PhRvL.118n7401W} which means we can parametrize the Moir\'{e} potential using only two parameters $(V, \psi)$ such that $V_1 = V e^{ i \psi}$.

The actual magnitude of the Moir\'{e} potential can be estimated using {\em ab initio} density functional theory. For this we use Quantum Espresso\cite{2017JPCM...29T5901G,2009JPCM...21M5502G} with a Coulomb cut-off\cite{Sohier:2017fl} to reproduce the two-dimensional nature of the heterostructures. We used a lattice structure where the interlayer distance, as measured by the $z$-distance between the W atoms, is $d_{W-W}=6.6$\AA, and the $z$-distance between the W and chalcogenides is 1.65\AA. The idea, following Refs.~\cite{Wu:2018ic,2018PhRvB..97c5306W,2017PhRvL.118n7401W}, is to calculate the energy of the {\em top of the valence band} in a small unit cell WS$_2$/WSe$_2$ bilayer where the top layer is shifted with a displacement ${\bf d}$. In the full Moir\'{e} unit cell, the Moir\'{e} potential follows the same energy dependence as the top of the valence band in the small unit cell.

\begin{figure*}[ht]
	\includegraphics[width=\textwidth]{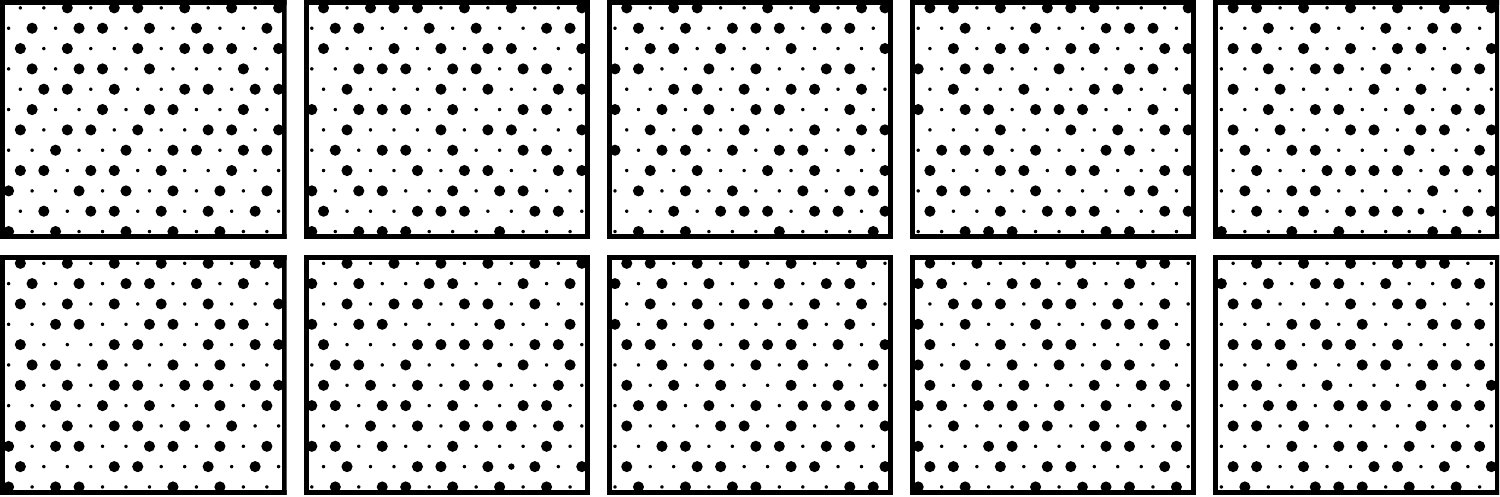}
	\caption{The Hartree code allows for many different amorphous charge configurations. Here we show 10 different such configurations for $V_0 = 30$ meV, at $T = 0.1$ meV, $n=1/2$ and $L=12$. }
	\label{Fig:Configs}
\end{figure*}
For aligned WS$_2$/WSe$_2$ bilayers we obtained $(V,\psi) = (7.7$ meV$, -106^\circ$), see Fig.~\ref{Fig:Moire}a. This corresponds to a Moir\'{e} potential peaked where the Se atoms are above the W atoms in the other layer, consistent with scanning tunnelling microscopy (STM) measurements.\cite{Li:2021cc} Note however, that the accuracy of the {\em magnitude} $V$ is debatable. Recently, STM measurements revealed a Moir\'{e} potential in excess of 100 meV,\cite{Shabani:2021fh} whereas theoretical work suggests a complete absence of a Moir\'{e} potential.\cite{Vitale:2021vg} Additionally, the magnitude is very sensitive to the precise interlayer distance (and can thus be tuned using pressure as was done in twisted bilayer graphene\cite{Yankowitz:2018tx}). Therefore, throughout this work, one should bear in mind that the provided values for the electronic bandwidth can be significantly different in real systems.

The potential Eq.~\eqref{Eq:MoirePotential} causes backfolding of the bandstructure Eq.~\eqref{Eq:ValenceBandTop} into the mini-Brillouin zone. At the edge of the mini-Brillouin zone the potential opens up a gap, leading to the formation of flat bands. The bandstructure of aligned WS$_2$/WSe$_2$ is shown in Fig.~\ref{Fig:Moire}b. Since we are dealing with a single band (per valley), a straightforward Fourier transform provides us with the tight-binding parameters. The nearest neighbor hopping is $t = 1.9$ meV, and the next-nearest neighbor hopping $t' = -0.4$ meV. Longer-ranged hopping parameters fall off exponentially and can be safely ignored.

Similarly, Wannierization\cite{Marzari:2012eu} using a projection onto a Gaussian wavepacket ansatz yields the Wannier orbitals which, in this case, reduces to approximately Gaussian around the W/Se stacking center with width $\sigma = 0.125 a_M$. This simple shape of the Wannier orbital allows us to directly extract the onsite repulsion $U$, and longer ranged Coulomb interaction parameters $V$. For this, we use the standard screened Coulomb potential $V_{\mathrm{sc}}(r) = \frac{e^2}{4 \pi \epsilon}\left( \frac{1}{r} - \frac{1}{\sqrt{r^2+ d^2}}\right)$ from having a screening layer at distance $d$ from the TMD heterobilayer. With a screening length of $d=20$ nm, this provides an onsite repulsion of $\epsilon U = 1.7$ eV and a nearest neighbor repulsion of $\epsilon V = 0.6$ eV. Even with quite large values for the dielectric constant the onsite repulsion is stronger than the bandwidth, and indeed even $V \gg t$. 
Because both the bandwidth (via the Moir\'{e} potential and interlayer coupling) and the interaction strength (via engineering of the dielectric environment) are tunable, in the remainder of this paper we vary the ratio $U/t$. Our choices are $U/t = 38.3, 19.1$ and $12.8$, corresponding roughly to dielectric constants of $\epsilon = 23.4, 46.8$ and $70.2$, respectively. For the strongest choice of $U/t$, the corresponding nearest neighbor repulsion becomes $V/t = 13$ -- large enough to induce nontrivial charge order. 

Note that the size of the Wannier orbital changes upon changing the moir\'e length. In the literature, both $\sigma \sim a_M$\cite{Wang:2020jv} and $\sigma \sim \sqrt{a_M}$\cite{Wu:2018ic} have been reported. The latter limit is valid in the limit where the moir\'e potential $V$ is larger than the kinetic energy associated with the reciprocal moir\'e lattice vector $\frac{\hbar^2 {\bf G}^2}{2m^*}$, whereas the linear dependence is valid when $V$ is small. For untwisted WS$_2$/WSe$_2$, we find $\frac{\hbar^2 {\bf G}^2}{2m^*}\approx 87$ meV whereas $V=7.7$ meV, clearly placing WS$_2$/WSe$_2$ in the regime where the Wannier orbital size is proportional to the moir\'e length, $\sigma \propto a_M$.

\section{APPENDIX B: Spinless Hartree methods}

The Hartree self-energy is defined as
$\Sigma^H ({\bf r}) =  
\sum_{{\bf r}'} V ({\bf r} - {\bf r}') \langle n_{{\bf r}} \rangle$
where the Coulomb interaction is approximated as $V({\bf r}) = \frac{V_0}{r} e^{(r-1)/d} $ where $r$ is measured in Moir\'{e} lattice constants, $d = 20$ nm the screening length and we choose $V_0 = 30$ meV. Including the Hartree self-energy, the total free energy is given by
\begin{eqnarray}
    F &=& - \frac{1}{2} \sum_{ij} \langle n_i \rangle V({\bf r}_i - {\bf r}_j) \langle n_j \rangle  
    \nonumber \\ && 
    - T \sum_{{\bf k}} \mathrm{Tr} \log \left[ \mathbf{1} + e^{ - ( \hat{h}({\bf k}) + \hat{\Sigma}^{H}) / T} \right] 
    \label{Eq:HartreeFreeEnergy}
\end{eqnarray}
which is iteratively minimized, starting from a random initial configuration at low $T = 0.1$ meV, on various large supercells containing multiple moir\'{e} unit cells. 

To obtain the free energy of commensurate charge ordering patterns (Fig.~2 of the main manuscript), we take small supercells containing 2, 3, 4 or 7 moir\'e unit cells. For these supercells there is a unique configuration that yields the lowest free energy. For example, for the curve corresponding to “1/3” order, we calculated the Hartree solution where the lattice symmetry is broken to a larger supercell containing 3 moiré unit cells. This means the charge density can be different on the three moiré unit cells contained in the supercell; let’s call them $n_1$, $n_2$, and $n_3$. The average filling is $n = (n_1 + n_2 + n_3)/3$. At exactly $n=1/3$ filling, the solution exists where $n_1 = 1$ and $n_2 = n_3 = 0$, which is shown in the side-panel of Fig.~2 of the main manuscript. At other fillings, the “1/3 order” curve corresponds to the lowest free energy possible given any choice of $n_1$, $n_2$, and $n_3$. 

To obtain the free energies at incommensurate charge orderings, we considered supercells containing $L \times L$ moir\'e unit cells with $L = 8, 12, 24$ and $36$. For each system size we started with a random initial configuration and found a (meta)stable mean field solution at zero $T$. After that, we increased temperature starting in each configuration to find the evolution of the charge order at finite temperature.

There are exponentially many different metastable configurations that can be found in this way. An indication of this is shown in Fig.~\ref{Fig:Configs} for $L=12$. The onsite Hartree energies act as input for the subsequent DMFT calculation, for which we calculated 100 metastable amorphous configurations at $L=12$.

\section{APPENDIX C: Dynamical Mean Field Theory methods}

Following the previous section, we know that the flat valence band
in TMD Moir\'{e} materials can be treated as an extended
Hubbard model on a triangular lattice

\begin{align}
H & =-\sum_{ij,\sigma}t_{ij}c_{i,\sigma}^{\dagger}c_{j,\sigma}+\sum_{i}Un_{i,\uparrow}n_{i,\downarrow}\nonumber \\
 & \ \ \ \ \ \ +\sum_{i,j\neq i,\sigma_{1},\sigma_{2}}V_{ij}n_{i,\sigma_{1}}n_{j,\sigma_{2}},\label{eq:Extended_Hubbard_Model}
\end{align}
where $c_{i,\sigma}^{\dagger},c_{i,\sigma}$ are the fermionic creation/annihilation
operators, $i,j$ are site labels to be summed over, $n_{i\sigma}\equiv c_{i\sigma}^{\dagger}c_{i\sigma}$
the density operator, $t_{ij}$ the hopping amplitude, $V_{ij}$ the
intersite long range Coulomb interaction strength and $U$ the onsite
Coulomb interaction strength. The hamiltonian Eq.\eqref{eq:Extended_Hubbard_Model}
differ from the single-band Hubbard model by considering also intersite
correlations, which can be reduced to a local form using Hartree theory
by replacing the density operator $n_{j,\sigma}$ with the spinless
expectation value $\left\langle n_{j}\right\rangle $
\begingroup\makeatletter\def\f@size{10}\check@mathfonts
\begin{align}
\sum_{i,j\neq i,\sigma_{1},\sigma_{2}}V_{ij}n_{i,\sigma_{1}}n_{j,\sigma_{2}} & \rightarrow\sum_{i,\sigma_{1}}n_{i,\sigma_{1}}\left(\sum_{j\neq i,\sigma_{2}}V_{ij}\left\langle n_{j}\right\rangle \right)\nonumber \\
 & =\sum_{i,\sigma}n_{i,\sigma}\varepsilon_{i},\label{eq:Hartree_approx}
\end{align}
\endgroup
where the effective site energy $\varepsilon_{i}$ is written as 

\begin{equation}
\varepsilon_{i}\equiv\sum_{j\neq i,\sigma_{2}}V_{ij}\left\langle n_{j}\right\rangle .
\end{equation}
By using the Hartree approximation Eq.~\eqref{eq:Hartree_approx},
all interactions in the extended Hubbard model Eq.~\eqref{eq:Extended_Hubbard_Model}
are now local

\begin{equation}
H_{\text{eff}}=-\sum_{ij,\sigma}t_{ij}c_{i,\sigma}^{\dagger}c_{j,\sigma}+\sum_{i,\sigma}n_{i,\sigma}\varepsilon_{i}+\sum_{i}Un_{i,\uparrow}n_{i,\downarrow}.\label{eq:eff_Hubbard_model}
\end{equation}

In the framework of single-site DMFT, Eq.~\eqref{eq:eff_Hubbard_model}
is solved first by identifying the corresponding Anderson impurity
problems then subject to DMFT self-consistency conditions on the bath
$\Delta\left(\omega\right)$. In below we will outline the general
single-site DMFT procedure to solve Eq.~\eqref{eq:eff_Hubbard_model}
for all commensurate fillings.

Except at $n=1$, the corresponding Anderson impurity problem consists of $m>1$
individual sites in the unit cell, depending on the Wigner crystal configuration in
Fig. \ref{Fig:Wigner} and hence we have

\begin{equation}
\Delta^{i=1,2,\dots,m}\left(\omega\right)\rightarrow\text{AIM solver}\rightarrow\Sigma_{i=1,2,\dots,m}\left(\omega\right),
\label{eq:DMFT_self_consistency}
\end{equation}
which is then subject to $m$ DMFT self-consistency equations

\begin{align}
\mathbf{\Delta}^{1}\left(\omega\right) & =\omega+\mu-\varepsilon_{1}-\Sigma_{1}-G_{11}^{-1}\left(\omega\right),\nonumber \\
\cdots & \cdots\nonumber \\
\mathbf{\Delta}^{m}\left(\omega\right) & =\omega+\mu-\varepsilon_{m}-\Sigma_{m}-G_{mm}^{-1}\left(\omega\right).\label{eq:self_consistency_general}
\end{align}
Specifically, in Eq.~\eqref{eq:DMFT_self_consistency}, we use real frequency Iterative Perturbation Theory (IPT) for arbitrary filling as the Anderson Impurity Solver(AIM solver)\cite{Kajueter:1996ew}. The term $G_{ii}\left(\omega\right)$ is computed from the local Green's
function $\mathbf{G}\left(\omega\right)$

\begin{align}
G_{ii}\left(\omega\right) & =\left[\dfrac{1}{N_{k}}\sum_{k}\mathbf{G}\left(\omega,\mathbf{k}\right)\right]_{ii},
\end{align}
where

\begin{align}
\mathbf{G}\left(\omega,\mathbf{k}\right) & =\dfrac{1}{\left(\omega+\mu+i\eta\right)\mathbf{I}-\mathbf{E}\left(\mathbf{k}\right)-\mathbf{\mathbf{\Sigma}}\left(\omega\right)},
\end{align}
$\eta$ is the broading term, ${\bf I}$ is the identity matrix, $\mathbf{E}\left(\mathbf{k}\right)$
is the dispersion matrix and $\mathbf{\Sigma}\left(\omega\right)$ is the diagonal
self-energy matrix

\begin{equation}
\mathbf{\Sigma}\left(\omega\right):=\begin{pmatrix}\Sigma_{1} & \cdots & 0\\
\vdots & \ddots & \vdots\\
0 & \cdots & \Sigma_{m}
\end{pmatrix}.
\end{equation}
A solution is obtained iteratively by: 1. starting from initial ansatzes
$\Delta^{i}\left(\omega\right)$, 2. solving the corresponding Anderson
impurity problems to obtain $\Sigma_{i}$, 3. use self-consistency
condition Eq.~\eqref{eq:self_consistency_general} to calculate $\Delta^{i}\left(\omega\right)$
for the next iteration.

Specifically for the triangular lattice at $1/3$ filling in this paper, the exact form of the band structure is the following:

\begin{eqnarray}
\label{eq:eqn2}
\mathbf{E}\left(\mathbf{k},\epsilon_{A},\epsilon_{B}\right) &=&	\left[\begin{array}{ccc}
\epsilon_{A} & 0 & 0\\
0 & \epsilon_{A} & 0\\
0 & 0 & \epsilon_{B}
\end{array}\right]+\mbox{\boldmath$\epsilon$}\left(\mathbf{k}\right) \nonumber \\
&=&
\left[\begin{array}{ccc}
\epsilon_{A} & E_{12} & E_{13}\\
E^{*}_{12} & \epsilon_{A} & E_{23}\\
E^{*}_{13} & E^{*}_{23} & \epsilon_{B}
\end{array}
\right]
\end{eqnarray}
where
\begin{eqnarray}
\label{eq:eqn3}
E_{12}=-t\left(e^{ik\delta_{ab}^{\left(1\right)}}+e^{ik\delta_{ab}^{\left(2\right)}}+e^{ik\delta_{ab}^{\left(3\right)}}\right);\nonumber\\
E_{13}=-t\left(e^{ik\delta_{ac}^{\left(1\right)}}+e^{ik\delta_{ac}^{\left(2\right)}}+e^{ik\delta_{ac}^{\left(3\right)}}\right);\\
E_{23}=-t\left(e^{ik\delta_{bc}^{\left(1\right)}}+e^{ik\delta_{bc}^{\left(2\right)}}+e^{ik\delta_{bc}^{\left(3\right)}}\right);\nonumber
\end{eqnarray}
and
\begin{eqnarray}
\label{eq:eqn4}
\delta_{ac}^{\left(1\right)}	&=&\left(a,0\right), 
  \delta_{ac}^{\left(2\right)}	=\left(-\frac{a}{2},\frac{\sqrt{3}a}{2}\right), \nonumber \\ 
\delta_{ac}^{\left(3\right)}	&=&\left(-\frac{a}{2},-\frac{\sqrt{3}a}{2}\right), 
\delta_{ab}^{\left(1\right)}=\delta_{bc}^{\left(1\right)}	=\left(-a,0\right),  \nonumber \\ 
\delta_{ab}^{\left(2\right)}&=&\delta_{bc}^{\left(2\right)}	=\left(\frac{a}{2},\frac{\sqrt{3}a}{2}\right), 
\delta_{ab}^{\left(3\right)}=\delta_{bc}^{\left(3\right)}	=\left(\frac{a}{2},-\frac{\sqrt{3}a}{2}\right),
\end{eqnarray}
and $a$ is the lattice spacing.
\begin{figure*}[ht]
	\includegraphics[width=0.4\textwidth]{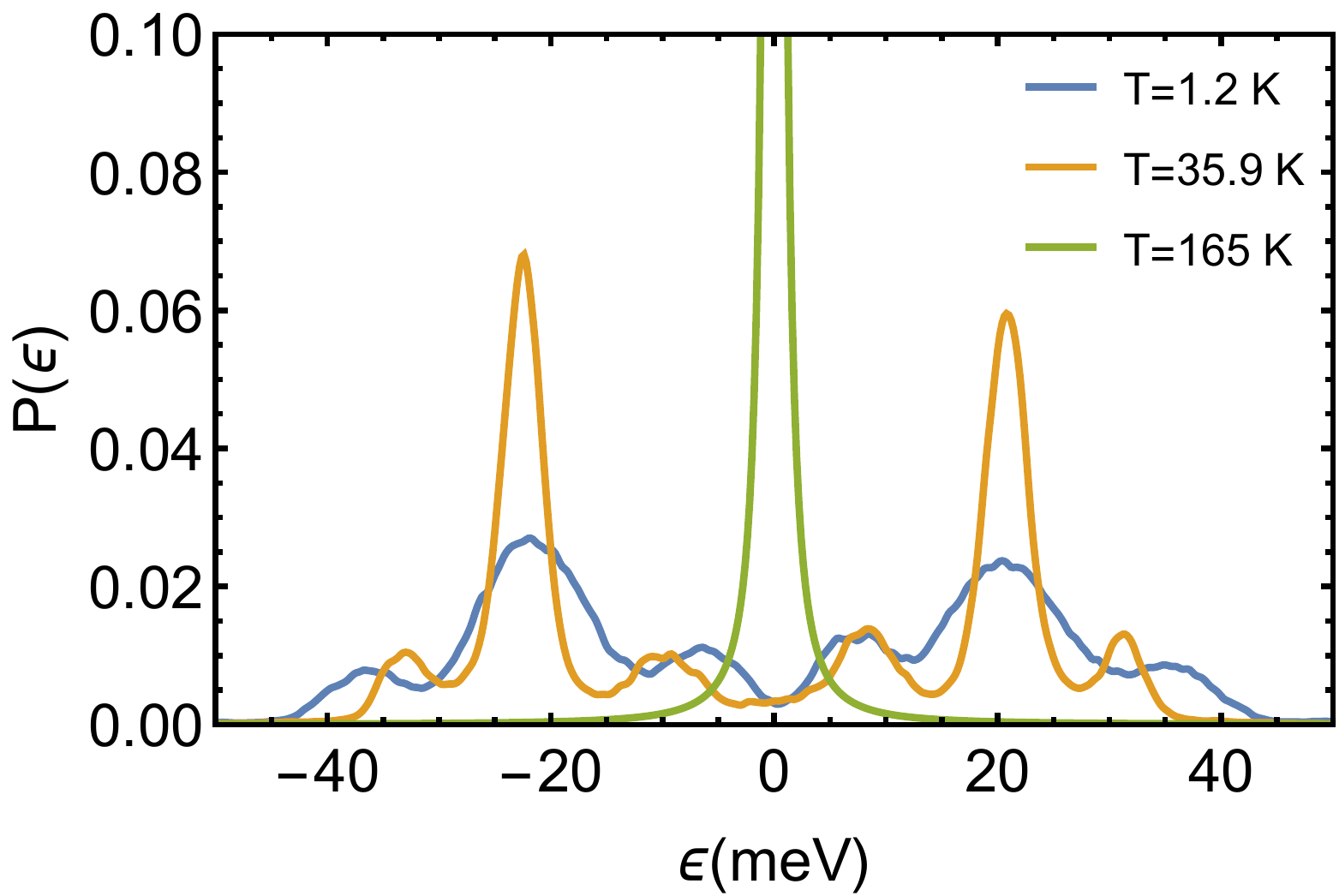}
	\caption{Histogram for amorphous state at $T=1.2K, 35.9K, 165K$.}
	\label{Fig:histogram}
\end{figure*}

In the case of an amorphous moire lattice, there is no periodicity
hence the system cannot be reduced to repeating unit cells (for $n=1/2$
see Fig.~\ref{Fig:Wigner}). Calculations indicate that
the Hartree site-energy is essentially randomly distributed and follow
a certain probability distribution $P\left(\varepsilon\right)$, shown in Fig.~\ref{Fig:histogram}. A
possible mean-field description of such a system is the Coherent Potential Approximation (CPA). In CPA,
spatial variation is disregarded, such that the impurity within the
DMFT framework is replaced by the average impurity, described by average
Green's function $\bar{G}\left(\omega\right)$
\begin{eqnarray}
\bar{G}\left(\omega\right)&=&\int d\varepsilon P\left(\varepsilon\right)G\left(\epsilon,\omega\right)\nonumber \\
&=&\int d\varepsilon\dfrac{P\left(\varepsilon\right)}{\omega+\mu-\varepsilon-\Delta\left(\omega\right)-\Sigma\left(\omega,\varepsilon\right)+i\eta},
\label{eq:average_green_function}
\end{eqnarray}
where $-\dfrac{1}{\pi}\Im{G\left(\epsilon,\omega\right)}$ gives the LDOS depending on the local Hartree site-energy. Eq.~\eqref{eq:average_green_function} leads to the effective self-energy

\begin{equation}
\bar{\Sigma}\left(\omega\right)=\omega+\mu-\bar{\varepsilon}-\Delta\left(\omega\right)-\bar{G}^{-1}\left(\omega\right)+i\eta.\label{eq:average_self_energy}
\end{equation}
The DMFT self-consistency equation is closed by
\begin{equation}
\Delta\left(\omega\right)=\omega+\mu-\bar{\varepsilon}-\bar{\Sigma}\left(\omega\right)-G_{\text{loc}}^{-1}\left(\omega\right).
\label{eq:hybridazation_function}
\end{equation}
where $G_{\text{loc}}\left(\omega\right)=\dfrac{1}{N_{\textbf{k}}}\sum_{\textbf{k}}\left(\omega+\mu-\bar{\epsilon}-\bar{\Sigma}\left(\omega\right)-E_{\textbf{k}}+i\eta\right)^{-1}$, and $E_{\textbf{k}}=-2t\left[\text{cos}\left(k_x\right)+2\text{cos}\left(\sqrt{3}/2k_y\right)\text{cos}\left(1/2k_y\right)\right]$.

To sum up, the DMFT+CPA recipe for amorphous system is as follows:

\begin{enumerate}
\item Input hybridization function $\Delta\left(\omega\right)$;

\item Solve $\Sigma\left(\omega,\varepsilon\right)$ for each Hartree site energy $\varepsilon$, using AIM impurity solver (Eq.~\eqref{eq:DMFT_self_consistency});

\item Compute average Green's function $\bar{G}\left(\omega\right)$ (Eq.~\eqref{eq:average_green_function});

\item Compute effective self-energy $\bar{\Sigma}\left(\omega\right)$ (Eq.~\eqref{eq:average_self_energy});

\item Find $\Delta\left(\omega\right)$ for the next iteration and repeat from 1 till converge (Eq.~\eqref{eq:hybridazation_function}).

\item The chemical potential $\mu$ is adjusted so that $n = 1/2$ with $10^{-3}$ accuracy.
\end{enumerate}

To get a good statistic, in step 2, we need to sample at least 100 Hartree site energies (solve for 100 impurity problems). Each DMFT loop takes 30 iterations on average, and on top of the DMFT loop, in step 6, the chemical potential $\mu$ needs to be adjusted $5\sim10$ times to keep $n$ fixed. So, to sum up, for one histogram $P(\epsilon)$, we need to solve roughly $100*30*5=15\text{K}$ impurity problems.

\begin{figure*}[ht]
	\includegraphics[width=16cm]{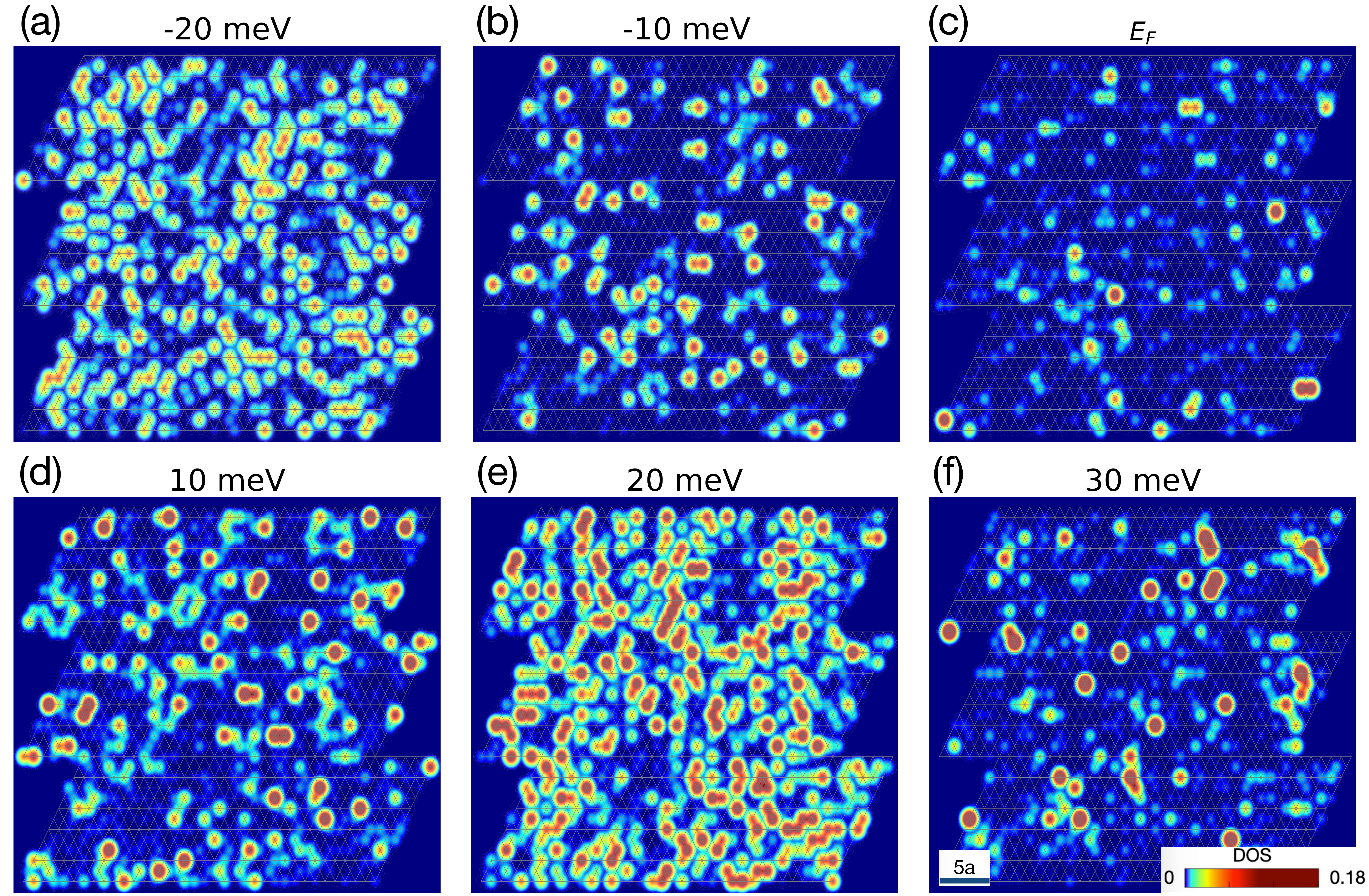}
	\caption{The spatial variations in the LDOS over an area of $36\times36$ unit cells, for the states below (a-b), at (c) and above (d-f) the Fermi level. The spectroscopic mapping use the same color scale for all panels. $T=1.2$K, $n=1/2$, $U/t=38.3$. }
	\label{Fig:moire_contour}
\end{figure*}

To obtain the LDOS mapping of large area shown in Fig.~\ref{Fig:Amorphous} and Fig.~\ref{Fig:moire_contour}, one firstly obtains how exactly the electrons freeze in an amorphous charge ordered pattern, in other words, the exact location $\left(x_i,y_i\right)$ of each Hartree site-energy $\epsilon_i$, via minimizing the total free energy on multiple Moir\'{e} unit cells. Then the LDOS for the states at energy $\omega$ is

\begin{equation}
\text{LDOS}\left(\omega\right)=\sum_i-\dfrac{1}{\pi}\Im{G\left(\epsilon_i,\omega\right)}\text{exp}\left[-\dfrac{\left(x-x_{i}\right)^2}{2\delta_{x}^2}-\dfrac{\left(y-y_{i}\right)^2}{2\delta_{y}^2}\right],
\end{equation}
where $\delta_x^2, \delta_y^2 $ are chosen to be $0.12 a_M$ , so that the width of the Gaussians is approximately the width of the Wannier functions, which is $0.125 a_M$.

\section{APPENDIX D: Phase diagram and transport properties around $n=1/3$}

For transport calculations, the DC conductivity is calculated via Kubo formula \cite{Economou}.:
\begin{eqnarray}
\label{eq:eqn13}
\sigma_{\mu\nu}&=&\frac{2e^{2}}{\pi\Omega}\int d\omega [-f^{'}\left(\omega\right)] \times\nonumber \\
&&\sum_{\mathbf{k}}\mbox{Tr}\left\{ \frac{ \partial\mathbf{E}{\left(\mathbf{k}\right)}}{\partial k_{\mu}}\Im\mathbf{G}\left(\omega,\mathbf{k}\right)\frac{\partial\mathbf{E}\left(\mathbf{k}\right)}{\partial k_{\nu}}\Im \mathbf{G}\left(\omega,\mathbf{k}\right) \right\}.
\end{eqnarray}
with
\begin{widetext}
\begingroup\makeatletter\def\f@size{11}\check@mathfonts
\begin{eqnarray}
\label{eq:eqn14}
\mathbf{v}_{\mu}\left(\mathbf{k}\right)&=&\frac{\partial\mathbf{E}\left(\mathbf{k}\right)}{\partial k_{\mu}} \nonumber \\
&=&
\left[\begin{array}{ccc}
0 & -it\sum_{n}\left(\delta_{ab}^{\left(n\right)}\right)_{\mu}e^{ik\delta_{ab}^{\left(n\right)}} & -it\sum_{n}\left(\delta_{ab}^{\left(n\right)}\right)_{\mu}e^{ik\delta_{ac}^{\left(n\right)}}\\
it\sum_{n}\left(\delta_{ab}^{\left(n\right)}\right)_{\mu}e^{-ik\delta_{ab}^{\left(n\right)}} & 0 & -it\sum_{n}\left(\delta_{ab}^{\left(n\right)}\right)_{\mu}e^{ik\delta_{bc}^{\left(n\right)}}\\
it\sum_{n}\left(\delta_{ab}^{\left(n\right)}\right)_{\mu}e^{-ik\delta_{ac}^{\left(n\right)}} & it\sum_{n}\left(\delta_{ab}^{\left(n\right)}\right)_{\mu}e^{-ik\delta_{bc}^{\left(n\right)}} & 0
\end{array}\right]
\end{eqnarray}
\endgroup
\end{widetext}
We only focus on the $xx$ component of the conductivity. For amorphous state, $\mathbf{E}\left(\mathbf{k}\right)$ is replaced by $\mathbf{\epsilon}\left(\mathbf{k}\right)$.

\begin{figure*}[ht]
	\includegraphics[width=12cm]{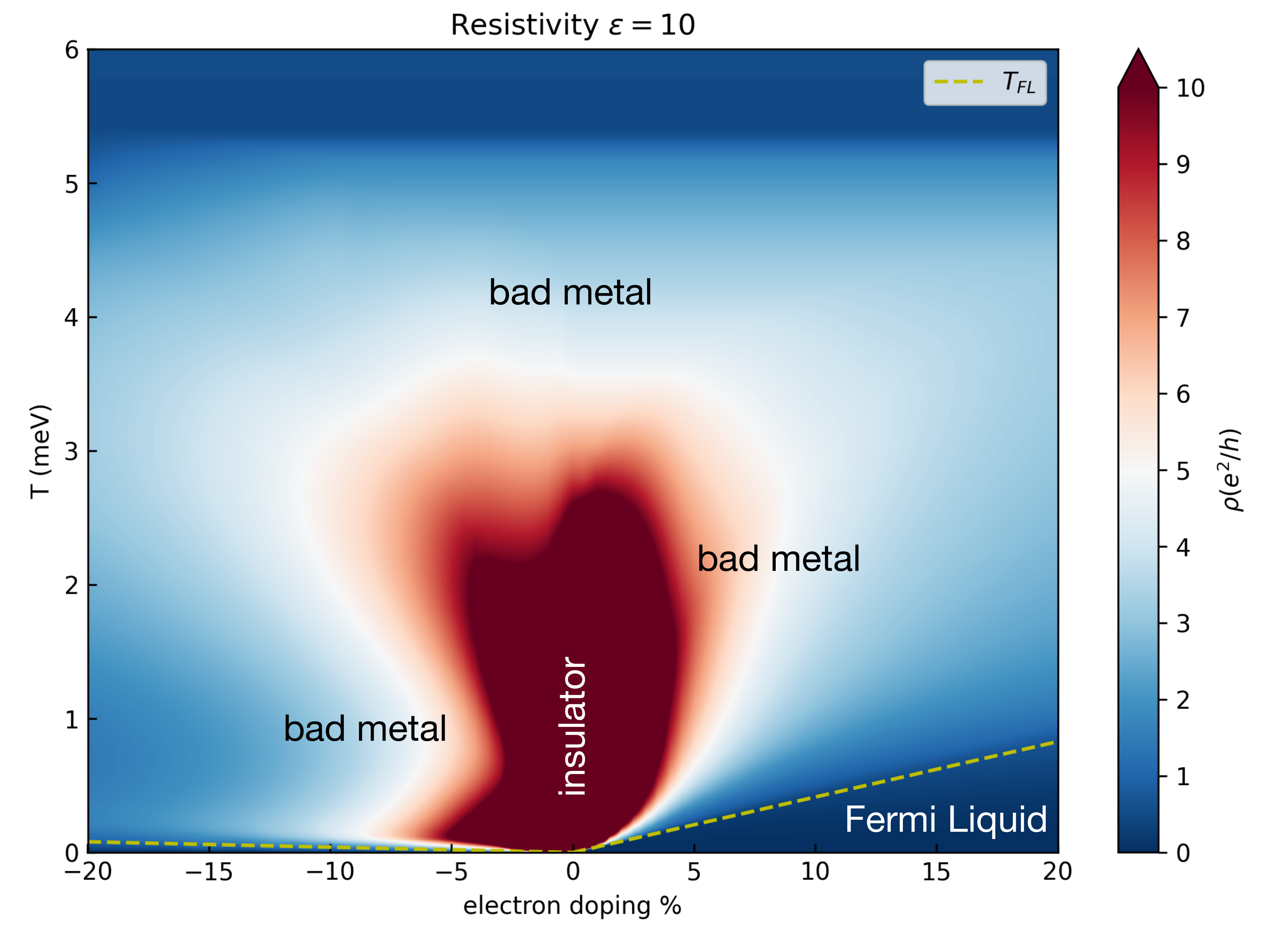}
	\caption{The phase diagram around the $n = 1/3$ Moir\'{e}-Mott Wigner frozen state in twisted TMDs. }
	\label{Fig:resistivity_contour}
\end{figure*}

In our results (Fig.~\ref{Fig:resistivity_contour}), we find that in the homogeneous case a Wigner-Mott insulating state around the commensurate filling at low temperature, while at high temperature charge-order is lost and the system becomes that of a bad metal with resistivity exceeding that of the Mott-Ioffe-Regel limit. We also investigate the effects of doping away from the commensurate filling, which results in a fermi-liquid at low temperatures and bad metal at higher temperatures. Exactly at $n = 1/3$ filling, the system becomes insulating due to a combination of charge order and Mott localization leading to a large charge-transfer gap. At temperatures where the charge order vanishes, linear resistivity is found indicating bad metal behavior. Electron or hole doping yield different phases due to the asymmetric nature of the charge-transfer gap. Upon electron doping, the low temperature phase is a heavy Fermi liquid with $T_{FL}$ vanishing linearly with doping. On the hole doped side, the Fermi liquid is pushed to extremely low temperatures.

\begin{figure*}[ht]
	\includegraphics[width=14cm]{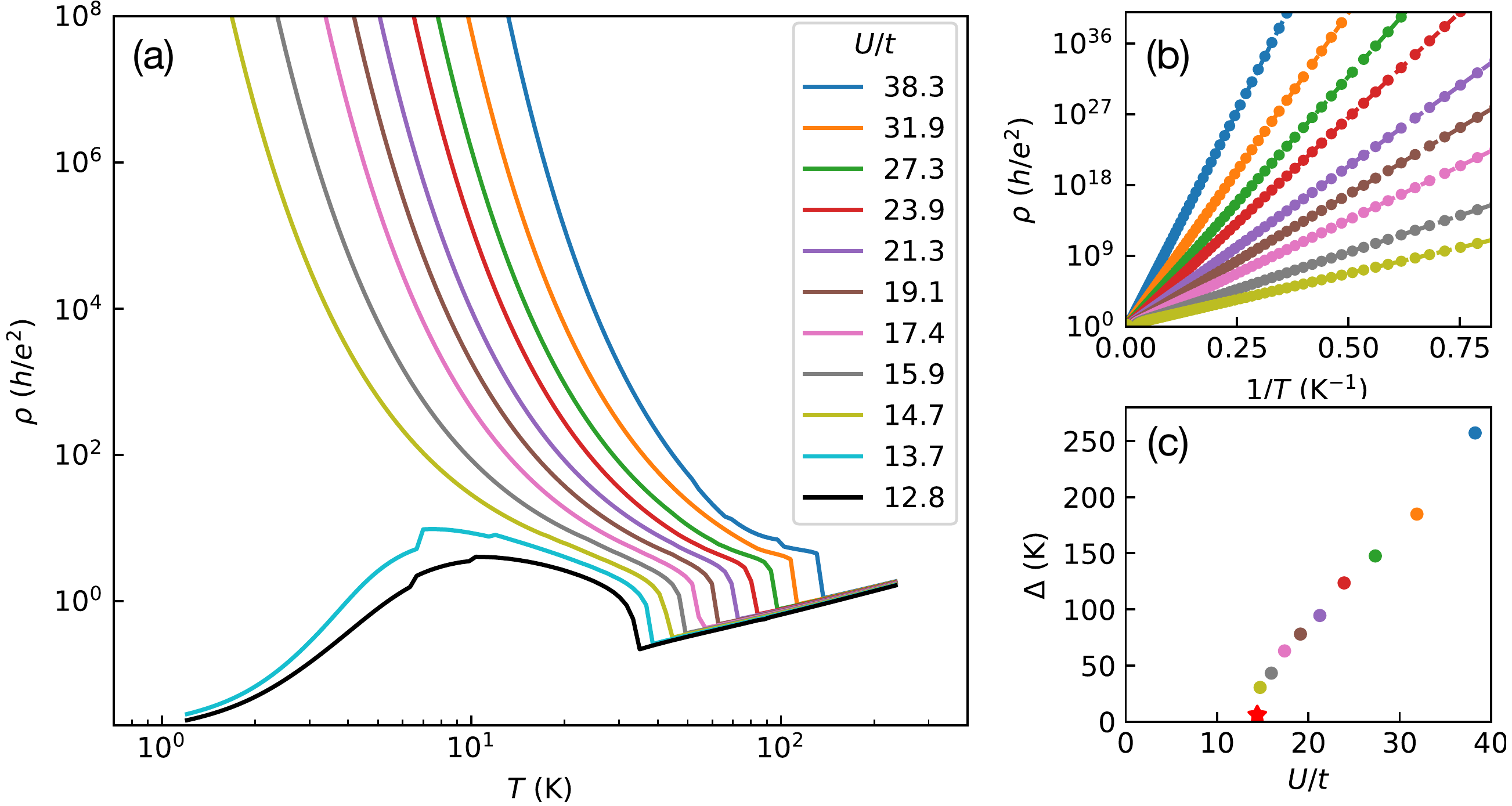}
	\caption{(a) DC resistivity curves for various $U$ at $n = 1/3$; (b) the Arrhenius plot for the insulating curves in (a); dashed lins are obtained from linear fitting; (c) the activation gap extracted from (b); the red star ping points  where metallic quasi-particles arise. }
	\label{Fig:arrhenuius}
\end{figure*}

Fig.~\ref{Fig:arrhenuius} shows the DC resistivity curves at $n = 1/3$ for various U across the transition. The MIT here has an essentially continuous character, similarly as for the Mott $f=1$ state. We also extract the activation gap from the Arrhenius plot. We can see the activation gap smoothly goes down but it extrapolate to zero at lower $U/t$ than where the quasi-particles arise (Red star in Fig.~\ref{Fig:arrhenuius}(c)), since it's a weakly first order transition. 

\section{APPENDIX E: Stability of the electron slush}

In this section we discuss whether the electron slush is a stable phase if we would include higher order corrections to our approximation scheme for the electron self-energy.

\begin{figure*}[ht]
        \includegraphics[height=5cm]{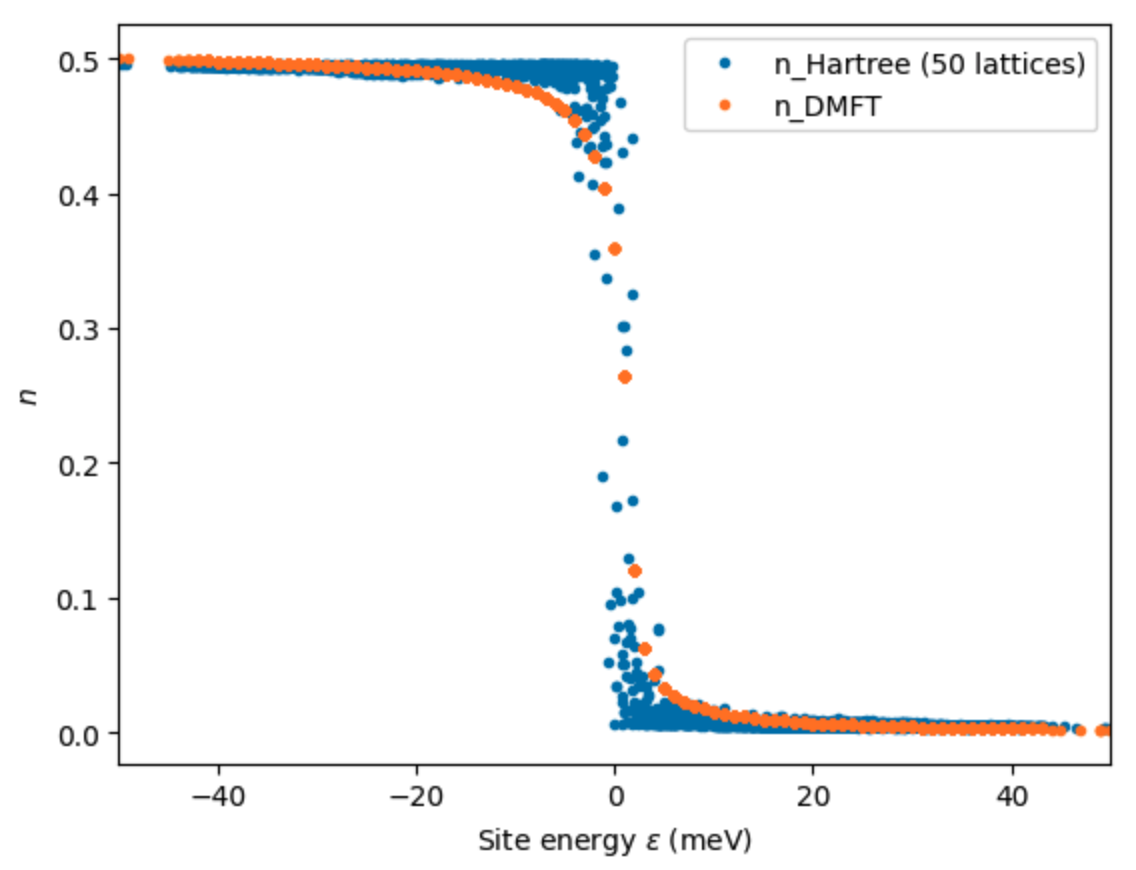}
	\caption{The occupation number of a collection of sites in the electron slush phase at $T=1.2$ K, based on either the Hartree self-energy only (in blue) or upon inclusion of the local self-energy calculated by DMFT (in orange).}
	\label{Fig:stabilitySlush}
\end{figure*}

Note that in the classical limit of very small $t$, the existence of amorphous metastable states is well established. This has been discussed in, for example, Ref. \cite{Rademaker:2018ku}. Such amorphous metastable states on a triangular lattice with Coulomb interactions have even been observed in organic compounds, see Refs. \cite{Kagawa:2013hz,Sato:2014cl}. 

This charge order is stable with respect to the introduction of a quantum hopping $t$. As such, introducing higher order corrections to the electron self-energy will inevitably lead to a reduction of the charge order but not to a complete suppression. Since charge order has been very clearly observed in WS$_2$/WSe$_2$, see Refs. \cite{Regan:2020fk,Xu:2020dx,Huang:2021io,Jin:2021es,LiWang:2021stmWigner,LiWang:2021stmLocal,Miao:2021exciton,Liu:2021exciton,Li:2021capacitance}, we are certainly in a range where nonlocal corrections to the electron self-energy affect quantitatively but not qualitatively the charge order.

What is left is the interplay between the developed charge order and the local onsite correlations as captured by DMFT. The question is whether the Hartree static mean field theory for the charges is affected by the non-static local self-energy.

To check this, we calculated the occupation number $n$ versus different Hartree site energies $\epsilon$ for the amorphous state for $T=1.2$ K, see Fig.~\ref{Fig:stabilitySlush}, and compare them with the occupation numbers {\em after} taking into account the DMFT solutions. The Hartree site energies were extracted from the distribution $P(\epsilon)$ as shown in Fig.~\ref{Fig:histogram}. Note that the average occupation number equals $\overline{n}=\int d\epsilon \, P(\epsilon) \, n(\epsilon)$ is 1/4 in the spinfull model. 

In Fig.~\ref{Fig:stabilitySlush} we see that while the DMFT rounds the charge occupations in a $\sim 10$ meV window, it does not qualitatively change the unequal site occupations characteristic of amorphous charge order. \YT{The randomess of $n_{Hartree}$ mainly comes from the randomness of a specific bath chosen in each calculation.} This is consistent with earlier works that fully self-consistently contained both onsite $U$ and nearest-neihbor $V$ on a square lattice\cite{Camjayi:2008jh}: the inclusion of large $U$ does not change the charge order transition.

Nevertheless, more quantitative accuracy is expected when one uses a technique that also includes the non-local dynamic self-energies, such as EDMFT.


\bibliographystyle{myapsrev}
\bibliography{library,triangularHubbard}

\end{document}